\documentclass[12pt]{article}
\pdfoutput=1

\usepackage{draft} 
\usepackage{hyperref}
\usepackage{graphicx,color,subfig}
\usepackage{cite}
\usepackage{mciteplus}
\usepackage{skak}
\usepackage{amsmath}
\DeclareFontFamily{OT1}{pzc}{}
\DeclareFontShape{OT1}{pzc}{m}{it}{<-> s * [1.10] pzcmi7t}{}
\DeclareMathAlphabet{\mathpzc}{OT1}{pzc}{m}{it}

\def\be#1\ee{\begin{align}#1\end{align}}
\newcommand\nn{\nonumber}
\def\({\left(}
\def\){\right)}
\begin{document}

\begin{titlepage}

	 \begin{flushright}
\hfill{\tt CALT-TH 2019-020, IPMU19-0086, PUPT-2586}
\end{flushright}

\begin{center}

\hfill \\
\hfill \\
\vskip 1cm

\title{Lightcone Modular Bootstrap and Pure Gravity
%A Twist on the Modular Bootstrap
%Please Mind the Gap
%Twist and Shout
}

\author{Nathan Benjamin$^{a}$, Hirosi Ooguri$^{b,c}$,  Shu-Heng Shao$^d$, and Yifan Wang$^e$
}

\address{
${}^a$Princeton Center for Theoretical Science, Princeton University, 
Princeton, NJ 08544, USA
\\
${}^b$Walter Burke Institute for Theoretical Physics,  \\
California Institute of Technology, Pasadena, CA 91125, USA
\\
${}^c$ Kavli Institute for the Physics and Mathematics of the Universe,  \\
University of Tokyo, Kashiwa, 277-8583, Japan
\\
${}^d$School of Natural Sciences, Institute for Advanced Study, 
Princeton, NJ 08540, USA
\\
${}^e$Joseph Henry Laboratories, Princeton University, 
Princeton, NJ 08544, USA
}

\email{nathanb@princeton.edu, ooguri@caltech.edu, \\ shao@ias.edu, yifanw@princeton.edu}

\end{center}

\vfill

\abstract{
We explore the  large spin spectrum in  two-dimensional conformal field theories with a finite twist gap, using the modular bootstrap in the lightcone limit. 
By recursively solving the modular crossing equations associated to different $PSL(2,\mathbb{Z})$ elements, we identify the universal contribution to the density of large spin states from the vacuum in the dual channel. 
Our result takes the form of a sum over $PSL(2,\mathbb{Z})$ elements, whose leading term generalizes the usual Cardy formula  to a wider regime.  
Rather curiously, the contribution to the density of states from the vacuum becomes negative in a specific limit, which can be canceled by that from a non-vacuum Virasoro primary whose twist is no bigger than $c-1\over16$.  
This suggests a new upper bound of $c-1\over 16$ on the twist gap in any $c>1$ compact, unitary conformal field theory with a vacuum, which would in particular imply that  pure AdS$_3$ gravity does not exist.  
We confirm this negative density of states in the pure gravity partition function by Maloney, Witten, and Keller. We generalize our discussion to theories with $\mathcal{N}=(1,1)$ supersymmetry, and find similar results.
}

\vfill

\end{titlepage}

\eject

\tableofcontents

\unitlength = .8mm

\section{Introduction}

Despite  progress in the classification program of rational conformal field theories, we have shockingly little understanding of the general landscape of 2d conformal field theories (CFTs).  
For example, there is no explicit construction of any 2d compact, unitary CFT with central charge $c>1$ and no Virasoro conserved currents of any spin \cite{Collier:2016cls}.  Such CFTs are expected to be generic, and our ignorance of them clearly shows the limitation in our understanding.

To go beyond the realm of rational CFTs, we will consider CFTs with a finite twist gap. The twist of an operator is defined as $ \Delta- |j| = 2\text{min}(h,\bar h)$.  Theories with conserved currents, such as rational CFTs, necessarily have a vanishing twist gap.\footnote{A priori, there could be theories without conserved currents, but with an accumulation of operators towards vanishing twist, and therefore have zero twist gap.}
  For this reason we will think of the twist gap  as a measure on how irrational a CFT is.  

In this paper, we will address the following two general questions for  CFTs with a finite twist gap:
\begin{enumerate}
\item Is there a universal Cardy-like growth for the large spin states?
\item At a fixed central charge $c$, how large can the twist gap  be?
\end{enumerate}
These two questions are tied together by modular invariance of the torus partition function.

We start with the first question.  
We generalize Cardy's argument for the asymptotic growth of states with large scaling dimensions \cite{Cardy:1986ie}. 
More specifically, we analytically continue the torus moduli $\tau , \bar \tau$ to two independent complex variables, and consider the limit $\text{Im}(\tau )\to0$ while keeping $\bar \tau$ fixed. 
This is similar to the lightcone limit studied in the conformal bootstrap of four-point functions \cite{Fitzpatrick:2012yx,Komargodski:2012ek}, but now applied to the modular bootstrap program \cite{Hellerman:2009bu}.   
By solving the modular crossing equations in the lightcone limit with arbitrary rational real  parts of $\tau,\bar\tau$,\footnote{These are the $PSL(2,\bZ)$ images of the cusp at $\tau=i \infty$.} 
 we recursively identify the universal contribution from the Virasoro vacuum multiplet 
 to the density of large spin states for any $c>1$ CFT with a finite twist gap.  
Our formula generalizes the usual Cardy formula from the regime $h,\bar h\gg c$ to $ h\gg c$ but with $\bar h -{c-1\over24}>0$ finite.
The universal density of states takes the form of a sum over $PSL(2,\mathbb{Z})$ images,\footnote{More precisely, this is a sum over the coset $PSL(2,\mathbb{Z})/\Gamma_{\infty}$ where $\Gamma_{\infty}$ is the subgroup generated by $T:\tau \to \tau+1$ that stabilizes the cusp at $\tau=i\infty$.} whose leading term is the extended Cardy formula discussed recently in \cite{Kusuki:2018wpa, Collier:2018exn, Kusuki:2019gjs}.    
The density of states depends on the number-theoretic properties of the spin $j= h-\bar h$ and is in particular non-analytic in $j$.

Now we turn to the second question. 
Rather curiously, our density of states from the vacuum contribution becomes negative in the double limit 
where $ j \to \infty$ and $\bar h- {c-1\over24} \to0$.  
Such negative density of states of course should not be present in a physical, unitary CFT.  
This negativity can be canceled by the contribution from a non-vacuum primary operator of twist $\Delta- |j|$ at or below $c-1\over 16$ in the dual channel. 
We are therefore led to the following tentative conclusion: any compact, unitary CFT with a $PSL(2,\bC)$ invariant vacuum must  have a twist gap of at most $c-1\over16$.  
Our argument is not yet rigorous, and we will discuss the gaps to complete the proof. If true, our result improves the earlier $c-1\over12$ bound on the twist gap by T. Hartman and \cite{Collier:2016cls}.\footnote{In \cite{Collier:2016cls} this argument was credited to Tom Hartman.}

Via the holographic correspondence \cite{Maldacena:1997re}, our result has interesting implications on pure Einstein gravity as a quantum gravity theory in AdS$_3$.  
In the strictest sense, pure AdS$_3$ gravity is dual to a  2d large $c$, unitary CFT where all non-vacuum Virasoro primary operators have $h,\bar h\ge {c-1\over24}$ and are interpreted as Ba\~{n}ados-Teitelboim-Zanelli (BTZ) black holes.  
The new twist gap bound $c-1\over16$ suggested by our argument would imply that pure AdS$_3$ gravity does not exist.\footnote{Recently, the closest theory to pure gravity in AdS$_2$, the Jackiw-Teitelboim theory, has been shown to be dual to a random matrix model, rather than a single quantum system with a definite Hamiltonian \cite{Saad:2019lba}.  
Furthermore, a pure AdS gravity theory, if exists, would have been a counterexample to the swampland conjecture in \cite{Ooguri:2016pdq}. 
The current paper provides another piece of evidence that pure gravity in AdS$_3$ might not be dual to a single unitary 2d CFT.} 
Indeed, we will check explicitly that the pure gravity partition function computed by Maloney-Witten-Keller \cite{Maloney:2007ud,Keller:2014xba} agrees with our formula in the specific double limit mentioned above.  
The sum over the $PSL(2,\mathbb{Z})$ elements in our formula is identified as a sum over geometries in AdS$_3$.\footnote{A similar sum over $PSL(2,\mathbb{Z})$ images was originally  interpreted as a sum over gravitational saddle points in \cite{Dijkgraaf:2000fq} for elliptic genera. Each term in the $PSL(2,\mathbb{Z})$ sum corresponds to a different AdS$_3$ geometry discussed in \cite{Maldacena:1998bw}.}
In particular, we confirm that the pure gravity partition function has an identical negative density of states in this limit.  
 This gives another interpretation of our result: while the pure gravity partition function of  \cite{Maloney:2007ud,Keller:2014xba}  is unphysical in various different ways, it approximates the universal density of large spin states dictated by the vacuum state in the dual channel of the modular crossing equation.   In other words, the pure gravity partition function of Maloney-Witten-Keller is the analog of double-twist operators in $d>2$ \cite{Fitzpatrick:2012yx,Komargodski:2012ek}, or of ``Virasoro Mean Field Theory" in 2d \cite{Collier:2018exn} for the modular bootstrap (see also \cite{Maloney:2016kee}).

The paper is organized as follows.   In Section \ref{sec:HCLY} we review the argument by T. Hartman and  \cite{Collier:2016cls} for the $c-1\over12$ bound on the twist gap.  In Section \ref{sec:CardyPlus},  the extended Cardy formula for the density of large spin states is reviewed.  In Section \ref{sec:turnonreal}, we generalize the extended Cardy formula to include subleading corrections by solving recursively the  crossing equations associated to general elements of  $PSL(2,\mathbb{Z})$.  In Section \ref{sec:positive}, the implications of this universal density of large spin states are discussed, which suggest that the twist gap in any compact, unitary $c>1$ CFT can be at most $c-1\over16$. 
 Section \ref{sec:pure} discusses the interpretation of our result in relation to the pure AdS$_3$ gravity partition function.   
 In Section \ref{sec:N=1}, we discuss the ${\cal N}=(1,1)$ supersymmetric generalization. 
Appendices \ref{sec:sumest} and  \ref{sec:plugin} describe some technical steps needed in solving the crossing equations.   
Appendix \ref{sec:accumulation} discusses some subtleties present when there is an accumulation of operators in twist.  
In Appendix \ref{sec:modularkernel}, we record the modular crossing kernels for more general elements of $PSL(2,\mathbb{Z})$.

\section{Warm-Up: The $c-1\over 12$ Twist Gap}
\label{sec:HCLY}

In this section we review an argument by T. Hartman  and  \cite{Collier:2016cls} showing that the twist gap in any compact unitary 2d CFT has to be no larger than $c-1\over12$.  This argument has been generalized from the Virasoro algebra to the ${\cal W}_N$ algebra in \cite{Afkhami-Jeddi:2017idc}.

Consider the partition function $Z(q,\bar q)$ of a 2d CFT on a torus with complex structure moduli $q=\exp(2\pi i \tau),~\bar q= \exp(-2\pi i \bar \tau)$.   
We will analytically continue so that $\tau$ and $\bar\tau$ are two independent complex variables. 
Let us parametrize the torus moduli as
\ie
\tau = i { \beta\over 2\pi}\,,~~~~\bar\tau = -  i{\overline\beta\over 2\pi} \,,
\fe
and take $\beta,\overline\beta$ to be independent positive numbers. 
The physical interpretation of this analytic continuation of $\tau,\bar\tau$ to two independent imaginary values is the following.  The torus partition function can be interpreted as the twist-field four-point function in the symmetric product of two identical copies of the original CFT. Then taking $\tau,\bar\tau$ independently to be purely imaginary corresponds to the Lorentzian regime of the twist-field four-point function.  
Later we will take $\beta\to0$ while keeping $\overline\beta$ fixed, which is the lightcone limit from the twist-field four-point function point of view.  We have
\ie
q=\exp\left(-\beta\right)\,,~~~~\bar q  = \exp\left( -\overline\beta  \right) \,.
\fe
Their modular $S$ transforms will be denoted with a prime:
\ie
q'  = \exp \left( -{4\pi^2\over\beta} \right)\,,~~~~~\bar q'  = \exp \left( -{4\pi^2 \over\overline\beta} \right) \,.
\fe

The torus partition function can be expanded in Virasoro characters.  
For a 2d unitary CFT with $c>1$, the possible modules of the Virasoro algebra are the degenerate module, {\it i.e.} the vacuum module $h=0$, and a continuous family of non-degenerate modules labeled by a positive conformal weight $h>0$.  
Their Virasoro characters are given by
\ie\label{vcharacter}
\chi_0(q)=(1-q){q^{-{c-1\over 24}}\over \eta(q)},~~~~\chi_{h>0}(q)={q^{h-{c-1\over 24}}\over \eta(q)}\,.
\fe
Combining the left with the right, there are three kinds of Virasoro primaries:
\ie\label{virch}
\text{(vacuum)} \quad & \chi_0(q){ \chi}_0(\bar q)\,,\\
\text{(conserved current)} \quad &\chi_0(q){\chi}_{\bar h>0}(\bar q), \quad \chi_{h>0}(q){ \chi}_0(\bar q)\,,
\\
\text{(non-degenerate)} \quad & \chi_{h> 0}(q){ \chi}_{\bar h> 0}(\bar q)\,.
\fe

Consider an operator with conformal weights ($h,\bar h)$. 
The scaling dimension $\Delta$ and the spin $j$ are defined as
\ie
\Delta=  h+\bar h \,,~~~~ j  =h-\bar h\,.
\fe
The twist of an operator is defined as $ \Delta- |j |$, and we will denote half of the twist as $t$:
\ie
t \equiv  \text{min}(h,\bar h) = {\Delta-  |j|\over2} \,.
\fe
We would like to study 2d CFTs with finite twist gap $2t_{\rm gap}>0$.  In particular, this implies that there is no conserved current in the theory. 
Under the finite twist gap assumption, 
the torus partition function can  be expanded as
\ie\label{LHS1}
Z(q,\bar q)   &= \chi_0(q) \chi_0(\bar q) + \sum_{h,\bar h\geq t_{\rm gap}} n_{h,\bar h} \, \chi_h(q) \bar\chi_{\bar h}(\bar q)\\
&= \sum_{j\in\mathbb{Z}}  \int_0^\infty dt  \, \rho_j (t) \, \chi_h(q) \bar\chi_{\bar h}(\bar q)\,,
\fe
where $n_{h,\bar h}\in \mathbb{N}$ is the number of Virasoro primaries with conformal weights $(h,\bar h)$. 
In the second line we have introduced the density of Virasoro primaries $\rho_j(t)$, defined as  a discrete sum over delta functions:
\ie
\rho_j(t) \equiv \sum_{\cal O~\text{with spin $j$}}  \delta(t- t_{\cal O} ) \,,
\label{phyden}
\fe
where the sum is over the Virasoro primaries  in the spectrum with spin $j$, and $t_{\mathcal{O}}$ refers to $\text{min}\(h_{\mathcal{O}},\bar h_{\mathcal{O}}\)$.  In what follows we will refer to $\rho_j(t)$ as the ``density of states" even though it is really the density of Virasoro primaries.

Using modular invariance $Z(q,\bar q) = Z(q',\bar q')$, we can rewrite the partition function in the dual channel:
\ie\label{RHS1}
Z(q',\bar q' )& = \sum_{j'\in\mathbb{Z}} \int_0^\infty dt' \,  \rho_{j'}(t' ) \, \chi_{h'}(q') \bar\chi_{\bar h'}(\bar q')\\
&= {\exp\left({  {4\pi^2 \over \beta}  {c-1\over24} } \right)  \over \eta( q')}
\left[
(1-e^{-{4\pi^2 / \beta}  } ) \chi_0(\bar q' )
  +  
\sum_{j\in\mathbb{Z}}\int_{t_{\rm gap} }^\infty dt' \,   \rho_{j'}(t') \, e^{-{4\pi^2  h' / \beta}}
\chi_{\bar h'}(\bar q' ) 
  \right]
\fe

Up to this point all the equations are exact with no approximation. We now equate \eqref{LHS1} with \eqref{RHS1} and take the $\beta\to0$ limit:
\ie\label{preS}
&
\text{vac}  +\sum_{j\in\bZ}\int_{t_{\rm gap} }^\infty dt\,  \rho_j(t) \, e^{-\beta( h-{c-1\over24})}e^{-\overline\beta (\bar h-{c-1\over24})} \\
 =& \sqrt{4\pi^2\over \beta\overline\beta}\,
  e^{ {4\pi^2\over \beta} {c-1\over24}  }  
   e^{ {4\pi^2\over \overline\beta} {c-1\over24}  }\,\left[ 
 (1-e^{-{ 4\pi^2\over\beta}}) (1-e^{- {4\pi^2\over\overline\beta}})
 +\cdots\right]
\fe
where vac $\equiv (1-q)(1-\bar q)q^{-{c-1\over24}}\bar q^{-{c-1\over24}}$.  We have used  $\eta(q') =\sqrt{\beta\over 2\pi}\, \eta( q)  $. 
The $\cdots$ are contributions from the non-vacuum operators in the cross channel.
As we take $\beta\to0$ (but keep $\overline\beta$ finite), the divergence on the RHS has to be reproduced by an infinite number of states on the LHS.

Let us further simplify the LHS of \eqref{preS} in the $\beta\to0$ limit.  First we can drop the vacuum term since any individual term does not give a divergence as $\beta\to0$.  We then write the LHS as
\ie\label{writespins}
&\sum_{j=0 }^\infty e^{ -\beta j } \int_{t_{\rm gap}}^\infty d\bar h \, \rho_j (\bar h) \exp\left[ -(\beta+\overline\beta) (\bar h-{c-1\over24}) \right]\\
+&\sum_{j=-\infty}^{-1} e^{\overline \beta j } \int_{t_{\rm gap}}^\infty dh \, \rho_{j} ( h) \exp\left[ -(\beta+\overline\beta) (h-{c-1\over24}) \right]  \,.
\fe
Note that $t=\bar h$ if $j\ge 0$ and $t= h$ if $j<0$.  
In the $\beta\to0$ limit (while keeping $\overline\beta$ finite), the second term is finite, which can thus be ignored.  Moreover, we can replace $\beta+\overline\beta$ in the first term by $\overline\beta$ in this limit.  
Hence, \eqref{preS} in the $\beta\to0$ limit becomes
\ie\label{SConstraint}
&\sum_{j=0 }^\infty e^{ -\beta j } \int_{t_{\rm gap}}^\infty d\bar h \, \rho_j (\bar h) \exp\left[ -\overline\beta (\bar h-{c-1\over24}) \right]\\
=& {2\pi \over \sqrt{\beta\overline\beta}}\,
  e^{ {4\pi^2\over \beta} {c-1\over24}  }  
   e^{ {4\pi^2\over \overline\beta} {c-1\over24}  }\,
 (1-e^{-{ 4\pi^2\over\beta}}) (1-e^{- {4\pi^2\over\overline\beta}})
 +{\cal O}(e^{{4\pi^2 \over \beta}({c-1\over24} - t_{\rm gap})  })
\fe

Using \eqref{SConstraint}, we now prove the twist gap $2t_{\rm gap}$ cannot be larger than ${c-1\over12}$.  Let us assume otherwise, \emph{i.e.} $2t_{\rm gap} >{c-1\over12}$. We multiply both sides by $e^{\overline\beta (t_{\rm gap}  -{c-1\over24})}$. Then the LHS has a negative $\overline\beta$-derivative, but the $\overline\beta$-derivative of the RHS will eventually be positive for large enough $\overline\beta$ (while still keeping $\overline\beta \ll 1/\beta$) due to the exponential growth of the factor $e^{\overline\beta (t_{\rm gap}  -{c-1\over24})}$. We therefore arrive at a contradiction.

\section{Extended Cardy Formula}
\label{sec:CardyPlus}

In this section, we will review the derivation in \cite{Kusuki:2018wpa, Collier:2018exn, Kusuki:2019gjs} of the universal spectrum of large spin Virasoro primaries for all   $c>1$ 2d CFTs with nonzero twist gap, \emph{i.e.} $2 t_{\text{gap}}>0$. 
We will argue that the physical density of states $\rho_j(\bar h)$ in the large spin $j\gg c$ limit   is universally approximated by 
\begin{align}
\rho^0_{j,1}(\bar h) = &
{ e^{ {4\pi\sqrt{ (\bar h+j)\(\frac{c-1}{24}\)  }} }\over \sqrt{ \left( \bar h + j -\frac{c-1}{24} \right)  \left(\bar  h -\frac{c-1}{24} \right)}} \,
\theta(\bar h-{c-1\over24})\label{extcardy}
\\
\times &
\left[
\cosh\left(4\pi \sqrt{\left({c-1\over 24}\right)\left(\bar h-{c-1\over 24}\right)}\right)
-
\cosh\left(4\pi \sqrt{\left({c-25\over 24}\right) \left(\bar h-{c-1\over 24}\right)}\right)
\right]\,.\nn
\end{align}
We use the superscript 0 to remind the reader that, much as the usual Cardy formula, $\rho^0_{j,1}(t)$ is a continuous function of the twist $2t$ that at large spin approximates the physical density of states $\rho_j(t)$, which is  a sum of delta functions. 
The meaning of the subscript 1, on the other hand, will become clear in Section \ref{sec:turnonreal}. 
Here  $\theta(x)$ is the Heaviside step function that equals $1$ if $x>0$ and 0 otherwise. 
Note that the twist gap for this solution is $2t_{\rm gap} = {c-1\over 12}$. 
In the limit $j\gg c$ with $\bar h>{c-1\over24}$, this can be written more compactly as 
\ie
\rho_{j,1}^0(\bar h)=K_S(\bar h+j)K_S(\bar h) \,,~~~
\label{ec1}
\fe
where $K_S(\bar h)$ is the modular kernel for the $S$-transformation  \cite{Zamolodchikov:2001ah}
\begin{align}\label{KS}
&K_S(\bar h) \\
=&\sqrt{2\over \bar h-{c-1\over 24}}\left[
\cosh\left(4\pi \sqrt{\left({c-1\over 24}\right)\left(\bar h-{c-1\over 24}\right)}\right) -
\cosh\left(4\pi \sqrt{\left({c-25\over 24}\right) \left(\bar h-{c-1\over 24}\right)}\right)
\right]\,. \nn
\end{align}

The asymptotic growth \eqref{extcardy} generalizes the usual Cardy formula \cite{Cardy:1986ie}, which holds  without assuming  $2t_{\rm gap}>0$, beyond its regime of applicability $h, \bar h \gg c$.  
In the $\bar h\gg c$ limit of \eqref{extcardy}, $\rho_{j,1}^0(\bar h)$ reduces to the usual Cardy formula:
\ie
\rho^0_{j,1}(\bar h) \to 
{1\over 2\sqrt{ \left( h -\frac{c-1}{24} \right)  \left(\bar  h -\frac{c-1}{24} \right)}}\,
e^{{4\pi\sqrt{  \(\frac{c-1}{24}\)  h}}+{4\pi\sqrt{\(\frac{c-1}{24}\)  \bar h}} }\,,~~~h, \bar h \gg c\,.
\label{cardy}
\fe
For this reason, we will refer to \eqref{extcardy} as the \textit{extended} Cardy formula. 

Before we verify that \eqref{extcardy} is a solution to our crossing equation \eqref{SConstraint}, we first discuss the defining property of the modular kernel $K_S(\bar h)$.  It relates the vacuum Virasoro character in one channel to the non-degenerate Virasoro characters in the crossed channel:
\ie\label{KSdef}
\chi_0(\bar q')=\int^\infty_{c-1\over 24} d\bar h K_S(\bar h)\chi_{\bar h}(\bar q)\,,
\fe
or equivalently,
\ie\label{KSdef2}
e^{{4\pi ^2\over \overline\B}{c-1\over 24}}(1- e^{-{4\pi ^2\over \overline\B}})=\sqrt{\overline \B\over 2\pi}\int_{c-1\over 24}^\infty d\bar h K_S(\bar h)  e^{-\overline\B(\bar h -{c-1\over 24})}
\fe
and similarly for the holomorphic (left-moving) characters.  This equation will be crucial for our crossing solution.

The argument leading to the extended Cardy formula \eqref{extcardy} is similar to the original argument by Cardy, but now in the lightcone limit where $\beta\to0$ while $\overline\beta$ is held fixed. This leads to the crossing equation \eqref{SConstraint}, where the divergence on the RHS needs to be reproduced by a certain asymptotic growth of states  with large spin.  
Below we show that \eqref{extcardy} is indeed a solution to the crossing equation.  Plugging in the solution \eqref{extcardy} into \eqref{SConstraint}, the LHS becomes
\begin{align}
\sum_{j=0}^{\infty} e^{-\beta j}& \int_{c-1\over24}^\infty d\bar h 
K_S(\bar h+j) K_S(\bar h)
 \exp\left[ -\overline\beta (\bar h-{c-1\over24}) \right] \,.
\label{checkextcardy}
\end{align}
In Appendix~\ref{sec:sumest}, we show that in the small $\B$ limit, the sum over $j$ in the above equation can be approximated by an integral over $j$, up to a term that is $\beta$-independent, which can be absorbed in the error term of the RHS of \eqref{SConstraint}. 
Shifting integration variables  and using  \eqref{KSdef2}, we then reproduce the leading divergent terms in the $\beta\to0$ limit on the RHS of \eqref{SConstraint}. 

Therefore, \eqref{extcardy} is indeed a solution to the crossing equation,  and gives the universal density of large spin states in any compact unitary CFT with a finite twist gap, up to an error that grows slower than $\exp\left(4\pi \sqrt{ {c-1\over24}  j} \right)$ in the large spin limit.  
On the other hand, $\rho^0_{j,1}=0$ if $\bar h< {c-1\over24}$, meaning that there is no exponential growth in the large spin limit.

Let us comment on the corrections to  \eqref{extcardy} in the large spin limit.  
First, there are error terms from approximating the discrete spectrum by a continuous density of states.  
This error for the density of states $\rho(\Delta)$ that is insensitive to the spin has recently been quantified in \cite{Mukhametzhanov:2019pzy,Ganguly:2019ksp} (see also Appendix C of \cite{Das:2017vej}).  
Second, there are contributions to the density of states from the lowest twist, non-vacuum primary operators, corresponding to the last term in \eqref{SConstraint}. 
Lastly, there are contributions coming from the vacuum but for different elements of $PSL(2,\bZ)$.  The last two corrections will be discussed in later sections.

We    emphasize that the converse of our statement here is also true, which follows simply from running our argument backwards. Namely, if a 2d CFT has a large spin spectrum that satisfies \eqref{extcardy}, it is guaranteed to have a nonzero twist gap $2t_{\rm gap}>0$. It would be interesting if there is a holographic interpretation of \eqref{extcardy} for $\bar h$ not in the Cardy regime, in terms of the entropy of BTZ black holes. If so, it may suggest the theories holographically dual to Einstein gravity in AdS$_3$ generically have nonzero twist gap.

\section{$PSL(2,\mathbb{Z})$ Modular Crossing Equations}
\label{sec:turnonreal}

In this section we will repeat the analyses in Sections \ref{sec:HCLY} and \ref{sec:CardyPlus}, but with a more general $PSL(2,\mathbb Z)$ transformation, and we will find qualitatively new behavior.

\subsection{Crossing Equation}
We now repeat the previous analysis but instead take\footnote{Note that there is a slight generalization of this we can consider, in which $\text{Re}(\tau) \neq \text{Re}(\bar\tau)$. We will not pursue this generalization in the current paper.}
\begin{align}\label{ttb}
\tau &= i\frac{\beta}{2\pi} + \frac rs, ~~~~\bar\tau = - i\frac{\overline\beta}{2\pi} + \frac{r}{s}\,.
\end{align}
Here $r$ and $s$ are two coprime integers, with $s$ positive. 
As before, we consider the limit $\beta \rightarrow 0$ at fixed $\overline\beta$.  
The $q$-variables are
\begin{align}
q=\exp\left(-\beta + \frac{2\pi i r}{s}\right)\,,~~~~\bar q  = \exp\left( -\overline\beta - \frac{2\pi i r}s \right).
\label{eq:q2}
\end{align}
Let us act  on $\tau,\bar \tau$ with the modular transformation 
\be
\begin{pmatrix}
a & b \\ s & -r
\end{pmatrix} \in SL(2,\mathbb{Z})\,,~~~-ar-bs=1\,,
\label{eq:sl2z}
\ee
to get 
\begin{align}
\tau' &= \frac{2\pi i}{\beta s^2} + \frac{a}{s} \,,~~~~~~~~~~~~~~~~~~~~\bar\tau' =  -\frac{2\pi i}{s^2\overline\beta} + \frac{a}{s}\,,\nn\\
q' &= \exp{\(-\frac{4\pi^2}{s^2\beta} + 2\pi i \frac{a}{s} \)} \,,~~~~\bar q' = \exp{\(-\frac{4\pi^2}{s^2\overline\beta} - 2\pi i \frac{a}{s} \)}.
\end{align}

Now we use modular invariance of  the torus partition function to set  (\ref{LHS1}) and (\ref{RHS1}) equal with the new parametrizations of $q$ and $\bar q$ as above. 
Since each individual term in \eqref{LHS1} is not sufficient to reproduce the divergence in \eqref{RHS1} as we take $\beta\to0$, we can drop the isolated vacuum contribution in (\ref{LHS1}) and obtain:
\begin{align}
&\sum_{j\in\bZ} \int_{t_{\rm gap}}^\infty \, dt \rho_j(t)\, \exp\left[{-\beta(h-\frac{c-1}{24})}  -\overline\beta(\bar h-\frac{c-1}{24}) + 2\pi i \frac{r}{s} (h - \bar h)\right] 
  \label{eq:finaleq2}\\
=&\frac{2\pi}{s\sqrt{\beta\overline\beta}} e^{\frac{4\pi^2(c-1)}{24s^2\beta}}e^{\frac{4\pi^2(c-1)}{24s^2\overline\beta}}
(1-e^{-\frac{4\pi^2}{s^2\beta}+\frac{2\pi ia} s}) (1-e^{-\frac{4\pi^2}{s^2\overline\beta}-\frac{2\pi ia} s})
+ \mathcal{O}\(e^{\frac{4\pi^2 }{s^2 \beta}\(\frac{c-1}{24}-t_{\text{gap}}\)}\).\nn
\end{align}
From (\ref{eq:sl2z}), we see that $ar\equiv-1~(\text{mod}~s)$.  Moreover (\ref{eq:finaleq2}) is invariant under $a\to a+s$. 
We will henceforth write $a$ as the modular inverse of $-r$ mod $s$, \emph{i.e.} $a =- (r^{-1})_s$.

Following the same steps to \eqref{SConstraint}, we simplify the above crossing equation in the $\beta\to0$ limit to 
\begin{align}
&\sum_{j=0}^{\infty} \exp \left[ -\left( \beta - {2\pi i r\over s} \right)j\right]\int_{t_{\text{gap}}}^{\infty}  d\bar h \, \rho_{j}(\bar h) \exp\left[{-\overline\beta(\bar h-\frac{c-1}{24})}\right]  \label{eq:finalcrossing}\\
&= \frac{2\pi}{s\sqrt{\beta\overline\beta}} \,
e^{\frac{4\pi^2}{s^2\beta}  {c-1\over24}}  e^{\frac{4\pi^2}{s^2\overline\beta}{c-1\over24}}
\left(1-e^{-\frac{4\pi^2}{s^2\beta}-\frac{2\pi i \(r^{-1}\)_s} s}\right) 
\left(1-e^{-\frac{4\pi^2}{s^2\overline\beta}+\frac{2\pi i\(r^{-1}\)_s} s}\right)+ \mathcal{O}\(e^{\frac{4\pi^2 }{s^2 \beta}\(\frac{c-1}{24}-t_{\text{gap}}\)}\).\nn
\end{align}
Note that we reproduce \eqref{SConstraint} if we take $r=0, s=1$. This is our main equation for the modular crossing equations associated to more general $PSL(2,\bZ)$ elements.

\subsection{Solution to the Crossing Equations}
\label{sec:solC}
The universal density of states in \eqref{extcardy} does not successfully reproduce the RHS of (\ref{eq:finalcrossing}).
 In particular, if we plug in \eqref{extcardy} into \eqref{eq:finalcrossing}, the LHS in the $\beta\to0$ is finite and fails to provide the divergence  $\exp{\(\frac{4\pi^2}{s^2\beta} {c-1\over24}\)}$ on the RHS of the equation.
See Appendix \ref{sec:plugin} for detailed derivations. 
This implies that there must be other universal contributions to the density of large spin states that, while being subleading to \eqref{extcardy}, are responsible for solving the more general crossing equations \eqref{eq:finalcrossing}. 

Our solution to (\ref{eq:finalcrossing}) will involve the Kloosterman sum defined by:
\ie\label{kloosterman}
S(j, J ; s)  = \sum_{\substack{r:\,\text{gcd}(r,s)=1  \\ 0 \leq r<s} }  \exp\left( 2\pi i  \, {  rj + (r^{-1})_s J\over s}\right)\,,
\fe
which is sensitive to the number-theoretic properties of the integer spin $j$.
Some basic properties of the Kloosterman sum are $S(j,J;s) = S(-j,-J;s)$, $S(j,J ;s) = S(J,j;s)$, and $S(j+s,J ;s) = S(j,J+s;s) =S(j,J;s)$.  In addition,
\ie\label{eq:abprop}
\sum_{j=0}^{\text{lcm}(s,s')-1} S(j,J;s') e^{ -{2\pi i rj \over s}}  = s 
e^{2\pi i (r^{-1})_s J \over s }\delta_{s,s'}\,.
\fe

Finally we define the following functions:
\begin{align}
d_0(h, s) &= \sqrt{\frac{2}{s(h-\frac{c-1}{24})}}\cosh\(\frac{4\pi}{s}\sqrt{\(h-\frac{c-1}{24}\)\(\frac{c-1}{24}\)}\)\theta(h-\frac{c-1}{24}) \nn\\
d_1(h, s) &= \sqrt{\frac{2}{s(h-\frac{c-1}{24})}}\cosh\(\frac{4\pi}{s}\sqrt{\(h-\frac{c-1}{24}\)\(\frac{c-25}{24}\)}\)\theta(h-\frac{c-1}{24})\,,
\end{align}
and 
\begin{align}
\rho_{j,s}(h, \bar h) &= S(j,0;s)  d_0(h, s) d_0(\bar h,s) - S(j,-1;s) d_0(h,s) d_1(\bar h,s)\nn\\
&~~ - S(j,1;s)d_1(h,s) d_0(\bar h,s) +  S(j,0;s)d_1(h,s) d_1(\bar h,s)\,.
\end{align}
In the large $j$ limit, the continuous solution to (\ref{eq:finalcrossing}) is given by\footnote{As in Section \ref{sec:CardyPlus}, we use the superscript $0$ to emphasize this is a continuous function in the twist  which approximates the physical density of states $\rho_j(t)$.  By contrast, the latter is a sum of delta functions.}
\be
\rho_j^0(\bar h) = \sum_{s'=1}\rho_{j,s'}(\bar h +j , \bar h)\,,~~~~j\gg c\,.
\label{eq:rho0}
\ee
Importantly, the dependence on the spin $j$ is highly non-analytic and depends on the number-theoretic property through the Kloosterman sum \eqref{kloosterman}.  
The $s'=1$ term is the extended Cardy formula \eqref{extcardy}, which grows as $\exp\left(4\pi \sqrt{ {c-1\over24} j}\,\right)$ in the large spin limit. The higher $s'$ terms grow as $\exp \left( {4\pi \over s'} \sqrt{ {c-1\over24}  j}\, \right)$ and are subleading corrections to \eqref{extcardy}.

Let us comment on the sum in $s'$. 
The solution  presented above is designed to reproduce the divergence on the RHS of  the crossing equation \eqref{eq:finalcrossing} as $\beta\to0$.  
However, the divergence on the RHS is only present if $s \lesssim 1/\sqrt{\beta}$.  
The divergence, if present, controls the density of states whose spins are of the order $j\sim 1/\beta$ due to the suppression factor $e^{-\beta j}$ on the LHS of \eqref{eq:finalcrossing}.  
It follows that for a fixed large spin $j$, we can only trust the solution if $s \lesssim \sqrt{j}$, therefore the sum in \eqref{eq:rho0} should be truncated before order $\sqrt{j}$.

For a fixed  $s$,  we show in Appendix  \ref{sec:plugin} that the modular crossing equations labeled by $r$ with gcd$(r,s)=1$ are solved by the term $\rho_{j,s} (\bar h+j ,\bar h)$ in the sum in \eqref{eq:rho0}.  
In particular, we show that, to leading order in $\beta\to0$,  the terms with $s' \neq s$ do not contribute to the modular crossing equations for any $r$ at a fixed $s$.

Just as in Section \ref{sec:CardyPlus}, where the  density of states for the extended Cardy formula can be interpreted as a product of modular kernels for the $S$ transformation, the  density of states we derived in this section can also be interpreted as modular kernels for more general $PSL(2,\mathbb Z)$ transformations. In Appendix \ref{sec:modularkernel}, we present simplified expressions for some of these modular kernels.

\section{Twist Gap Revisited}
\label{sec:positive}

Let us examine the solution \eqref{eq:rho0} to the crossing equation. In the large spin limit $j\gg c$, the leading terms are \
\ie
\rho^0 _j (\bar h )& =  \rho_{j,1}( h ,\bar h)+ \rho_{j,2}( h ,\bar h)+{\cal O}(e^{ {4\pi\over3} \sqrt{ ({c-1\over24}) j  }})\\
=&  \left[  \, d_0 (h , 1)- d_1( h , 1)\, \right]\left[  \, d_0 (\bar h  , 1)- d_1(\bar h , 1)\, \right]\\
+&(-1)^j   \left[  \, d_0 (h  , 2)+d_1(h , 2)\, \right]\left[  \, d_0 ( \bar h  , 2)+d_1( \bar h , 2)\, \right]+{\cal O}(e^{ {4\pi\over3} \sqrt{ ({c-1\over24}) j  }})\,,
\fe
where we write $\bar h+j  =h$ above to avoid cluttering. Now we further focus on the following part of the spectrum in the large spin limit $j\gg c$:
\ie\label{negativity}
0<\bar h-  {c-1\over 24} < {1\over 8\pi^2} \exp\left(-2\pi \sqrt{j \(\frac{c-1}{24}\)} \right)\,,~~~~~j:~\text{odd}\,,
\fe
In this regime, the first term $\rho_{j,1}$ is smaller in magnitude than the second one $\rho_{j,2}$. This is due to the different $\bar h$ dependence in $\rho_{j,1}$ and $\rho_{j,2}$. As $\bar h$ approaches $\frac{c-1}{24}$ from above, $\rho_{j,1}$ scales as $\sqrt{\bar h - \frac{c-1}{24}}$, whereas $\rho_{j,2}$ scales as $\frac{1}{\sqrt{\bar h -\frac{c-1}{24}}}$. In the region \eqref{negativity}, the different $\bar h$ dependence is enough to overcome the larger exponential in spin that $\rho_{j,1}$ has over $\rho_{j,2}$. 
Moreover, $\rho_{j,s}$ for $s\geq2$ scales as $\frac{1}{\sqrt{\bar h -\frac{c-1}{24}}}$ in the same regime, so we do not get qualitatively new behavior by considering higher $s$. 

Since we take $j$ to be an \textit{odd} integer, our crossing solution $\rho^0_j(\bar h)$ is large and negative from the sign $(-1)^j$. In particular in the large $j$ limit,
\be
\int_{\frac{c-1}{24}}^{\frac{c-1}{24}+\frac{1}{8\pi^2}e^{-2\pi\sqrt{j\(\frac{c-1}{24}\)}}} d\bar h\,\rho^0_j(\bar h) \simeq -\frac{\sqrt{2} e^{\pi\sqrt j}}{3\pi \sqrt j}.
\ee

The density of states of a physical CFT receives correction to the solution $\rho^0_j(\bar h)$ from various sources.  One obvious correction comes from the lowest twist $2t_{\rm gap}$ operator in the dual channel. 
Let the conformal weights of this lowest twist operator be $(t_{\rm gap} , \bar h_{\rm gap})$.\footnote{Here we implicitly assume there are only a finite number of low twist operators. More precisely, we assume there exists an $\epsilon > 0$ such that are a finite number of primary operators with $h<t_{\text{gap}}+\epsilon$. This is not necessarily the case: there could be an accumulation point in the twist. In Appendix \ref{sec:accumulation} we will consider a slight generalization of the argument in this section where we account for this possibility, which suggests a weaker bound that $t_{\text{gap}} \leq \frac{15}{16}\(\frac{c-1}{24}\)$. } 
 Repeating the same argument in Section \ref{sec:CardyPlus}, this lowest twist operator contribute to the density of states by
\begin{align}\label{lowesttwistguy}
&\rho^{\rm non-vac}_j (\bar h)
=
{2\over \sqrt{ (h -{c-1\over24} )(\bar h -{c-1\over24} ) }     }  \\
&
\times \cosh \left[ 4\pi \sqrt{ \left( {c-1\over24}  - t_{\rm gap} \right)\left( h -{c-1\over24}\right) }\,\right]
 \cosh \left[ 4\pi \sqrt{ \left( {c-1\over24}  - \bar h_{\rm gap} \right)\left( \bar h -{c-1\over24}\right) }\,\right]\nn
\end{align}
in the large $j$ limit.  
Note that this contribution to the density of states is exponentially large in the large spin limit only if the twist $2t_{\rm gap}$ is below $c-1\over12$, which is necessarily the case as we reviewed in Section \ref{sec:HCLY}. 
This positive contribution from the lowest twist non-vacuum primary can only overcome the negativity of $\rho^0_j(\bar h)$ in the regime \eqref{negativity} if
\ie\label{newtwist}
t_{\rm gap} \leq {c-1\over32}\,.
\fe
This leads  to a tempting conclusion that all two-dimensional, unitary, $c>1$ CFTs with unique normalizable vacua have twist gap $2t_{\text{gap}}$ no greater than $\frac{c-1}{16}$.\footnote{In \cite{Maxfield:2019hdt}, a spin-dependent shift in the BTZ threshold is discussed, which can potentially be another way to cure the negative density of states.}

This is not yet a rigorous derivation as there are potentially other corrections to the density of states that might cure the negativity in $\rho^0_j(\bar h)$. Among other things, there are error terms from approximating the discrete spectrum by a continuous density of states as mentioned in Section \ref{sec:CardyPlus}.  It would be interesting to extend the analysis in \cite{Mukhametzhanov:2019pzy,Ganguly:2019ksp,Das:2017vej} to control the error in a rigorous way.  For a CFT with a finite twist gap (and hence necessarily irrational), there is generally no huge degeneracy at a given energy level, so we expect the error due to granularity to be much smaller than that of a rational CFT. 

An equivalent, but perhaps more intuitive, explanation of our argument is the following. Let us define 
\be
\rho_j^{\text{odd}}(\bar h) = \frac{1-(-1)^j}2 \rho_j(\bar h),
\ee
\emph{i.e.}, the density of Virasoro primaries of odd spin. We can obtain a crossing equation for $\rho_j^{\text{odd}}(\bar h)$ by combining \eqref{eq:finalcrossing} for $\frac rs=0$ and $\frac rs=\frac12$: 
\begin{align}
\sum_{j=0}^{\infty} &e^{-\beta j} \int_{t_{\rm gap}}^{\infty} d\bar{h} \rho_j^{\text{odd}}(\bar h) e^{-\overline\beta(\bar h - \frac{c-1}{24})} = \nn\\
&\frac{\pi}{\sqrt{\beta\overline\beta}} \left(e^{\frac{4\pi^2}{\beta}\frac{c-1}{24}} e^{\frac{4\pi^2}{\overline\beta}\frac{c-1}{24}}(1-e^{-\frac{4\pi^2}{\beta}})(1-e^{-\frac{4\pi^2}{\overline\beta}}) - \frac12  e^{\frac{\pi^2}{\beta}\frac{c-1}{24}} e^{\frac{\pi^2}{\overline\beta}\frac{c-1}{24}} (1+e^{-\frac{\pi^2}{\beta}})(1+e^{-\frac{\pi^2}{\overline\beta}})\right) \nn\\& ~~~~~~~~~~+\mathcal O\(e^{\frac{4\pi^2}{\beta}(\frac{c-1}{24}-t_{\rm gap})}\).
\label{eq:crossingodd}
\end{align}
If we take the inverse Laplace transform of the RHS of (\ref{eq:crossingodd}) we would obtain the first two terms of (\ref{eq:rho0}); a twist gap of $\frac{c-1}{16}$ or below would cure the negativity in this density.\footnote{Readers may notice that the RHS of \eqref{eq:crossingodd} without the error term can become negative for $\bar \beta$ of $\mathcal O(e^{3\pi^2\over \beta})$. However our lightcone bootstrap analysis requires taking $\beta\to 0$ first. Thus this apparent negativity in the canonical ensemble cannot be trusted. 
}

In \cite{Collier:2016cls}, the authors found a bootstrap upper bound on the twist gap $2t_{\rm gap}$ that is numerically close to the analytic bound $c-1\over12$ reviewed in Section \ref{sec:HCLY}. A simple partition function that saturates this bound comes from the $c\ge 1$ Liouville theory
\ie
Z_{\rm Liouville}  (q,\bar q)  \propto  { 1\over \tau_2^{1\over2} |\eta(q)|^2}\,.
\fe
Alternatively, we could consider the $c=1$ compact boson at any finite radius, but now viewed as a partition function for a $c>1$ theory by shifting the vacuum Casimir energy. In this new interpretation, there is no vacuum, while the original vacuum of the $c=1$ compact boson is now an $h=\bar h={c-1\over24}$ primary. A notable difference of this example from the Liouville case is that the spectrum of primaries is discrete and includes all spins.

At first sight, this seems to imply that one cannot lower the twist gap below $c-1\over12$, which is in tension with our suggested twist gap $c-1\over16$.  However,  recall that in our argument, it is crucial that there is a normalizable vacuum with $h=\bar h=0$ in the spectrum for us to perform the lightcone bootstrap.  Neither the Liouville partition function nor the shifted compact boson partition function contain a vacuum state; therefore they need not obey the constraints we derived.  Similarly in the numerical modular bootstrap analysis, it is difficult to impose the condition that there is a normalizable vacuum in the spectrum. We therefore predict that the functionals found in \cite{Collier:2016cls} would have zeros above the twist gap that ``coalesce". In other words, as the truncation order in derivative increases, the zeros become denser and denser rather than approach a fixed spectrum. 

Interestingly, unlike in \cite{Mazac:2016qev}, the limit as the truncation order goes to infinity does not produce a nontrivial ``extremal functional". In the previous paragraph, we argued that the limit would produce a functional that vanishes for all integer spins with twist at least $\frac{c-1}{12}$. However in Section \ref{sec:pure}, we will review a construction by \cite{Maloney:2007ud,Keller:2014xba} that inputs any single state with twist below $\frac{c-1}{12}$, and produces a modular invariant function by adding states all with twist at least $\frac{c-1}{12}$. Since the proposed extremal functional would vanish on the crossing equation for this modular invariant function, it must in addition vanish on all states with twist below $\frac{c-1}{12}$. Therefore the extremal functional approaches zero as the truncation order goes to infinity.

The twist  $2t = 2 \text{min}(h,\bar h)= {c-1\over 16}$ has also appeared in other contexts.  In \cite{Collier:2018exn}, the authors introduced the notion of ``Virasoro Mean Field Theory", defined as the inversion of the vacuum Virasoro block for the sphere four-point function.   While Virasoro Mean Field Theory by itself does not give a consistent four-point function of a physical theory, it approximates the large spin CFT data of any compact, unitary 2d CFT with nonzero twist gap.   For identical external operators with conformal weight $(h,\bar h)$, the authors show that the associated Virasoro Mean Field Theory spectrum is qualitatively different for $h> {c-1\over 32}$ versus $h<{c-1\over32}$: In the former case, the spectrum consists only of a continuum above $h>{c-1\over24}$, while in the latter case, there are in addition a discrete set of primaries.\footnote{Note that in terms of the Liouville momentum $\alpha(h)>0$, defined as $h= \alpha(Q-\alpha)$ with $c=1+6Q^2$, we have $2\alpha({c-1\over32})=  \alpha({c-1\over24})$.} 
Indeed, the four-point function of identical scalar primaries with scaling dimension $\Delta = {c-1\over 16}$ is special. 
The four-point sphere conformal block with external scalar scaling dimension $c-1\over16$ and internal scalar scaling dimension $c-1\over12$ is a simple power of $|z|, |1-z|$, and is self-crossing invariant by itself (see, for example, (3.15) of \cite{Esterlis:2016psv}).  
Furthermore, it has been shown in \cite{Collier:2017shs} that $c-1\over 16$ is the minimal external scalar scaling dimension for a four-point function with only internal scalar primaries.\footnote{If the external scaling dimension is greater or equal to $c-1\over12$, then such four-point functions with only scalar internal primaries are realized by Liouville theory. If the external scaling dimension is between $c-1\over16$ and $c-1\over12$, such four-point functions can be obtained by analytically continuing the Liouville four-point function \cite{Collier:2017shs}.} 
This value of the twist has also appeared in the discussion of the Renyi entropy after a local quench \cite{Kusuki:2018wpa,Kusuki:2019gjs}.

\section{Pure Gravity}\label{sec:pure}

The twist gap that we have proposed has very interesting implications for two-dimensional CFTs holographically dual to large-radius Einstein gravity in AdS$_3$. Recall that the classical BTZ black hole has mass, $ M$, and angular momentum, $j$, related to the CFT conformal dimensions $h$ and $\bar{h}$ via \cite{Strominger:1997eq}
\begin{align}
{M}= \frac{1}{\ell_{\text{AdS}}}\(h+\bar{h}-\frac{c}{12}\)\,,~~~~~~j  = h-\bar{h}
\end{align} 
in the large $c$ limit. In particular, classical BTZ black holes have $M\ell_{\text{AdS}} \geq |j|$, which implies $h, \bar{h} \geq \frac{c}{24} + \mathcal{O}(1)$. 

There have been attempts to formulate a ``pure" theory of quantum gravity in AdS$_3$ \cite{Maloney:2007ud,Keller:2014xba}. Via AdS/CFT, the strictest definition of a pure theory of gravity is a 2d unitary CFT at large $c$ where all non-vacuum Virasoro primary operators can be interpreted as BTZ black holes.\footnote{The  extremal partition functions of \cite{Witten:2007kt} are also often referred to as partition functions of pure gravity. However, due to their holomorphic factorization, the partition functions of \cite{Witten:2007kt} contain non-vacuum states with vanishing twist, that therefore cannot be interpreted as BTZ black holes.  In this section we thus will focus on the  partition function of \cite{Maloney:2007ud,Keller:2014xba}, which obeys the stricter definition of pure gravity.} 
The  bound \eqref{newtwist} on the twist gap  suggested by our argument would imply that no such CFT exists. 
In this section we check that the pure gravity partition function computed in \cite{Maloney:2007ud,Keller:2014xba} indeed has negative density of states in the regime \eqref{negativity}, confirming our general argument.

Our main result \eqref{eq:rho0} is the universal contribution from the vacuum character in the dual channel to the large spin density of states. 
Its expression is very reminiscent of the sum over geometries in the calculation of the partition function of pure AdS$_3$ gravity \cite{Maloney:2007ud,Keller:2014xba}.  This is not a coincidence.  
We will further show  that in the limit  $\bar h-{c-1\over24}\to0$ and $j\gg c$, the pure gravity partition function matches  our $\rho^0_j(\bar h)$ \eqref{eq:rho0}.  
We therefore reach an important conclusion: Even though the pure gravity partition function derived in \cite{Maloney:2007ud,Keller:2014xba}  has various unphysical properties, it is the universal contribution from the vacuum character to the density of large spin states in any CFT with a finite twist gap.

\subsection{Maloney-Witten-Keller Partition Function}

The Maloney-Witten-Keller  (MWK) partition function \cite{Maloney:2007ud,Keller:2014xba}  is computed by starting with the vacuum Virasoro character
\ie\label{vacchi}
\chi_0(q)\chi_0(\bar q) =  { 1\over |\eta(q)|^2}  \left( q^{ - {c-1\over24}  }\bar q^{  \, - {c-1\over24}  }  -q^{ - {c-1\over24} +1 }\bar q^{  \, - {c-1\over24}  } 
-q^{ - {c-1\over24}  }\bar q^{ \,- {c-1\over24} +1 } +q^{ - {c-1\over24} +1 }\bar q^{ \, - {c-1\over24}  + 1} \right)\,,
\fe  and adding its $PSL(2,\mathbb{Z})$ images.  The sum is divergent, and a certain regularization is required to make the answer finite.  
The MWK partition function has the following features:
\begin{itemize}
\item It has a unique vacuum, and all other Virasoro primaries have $h,\bar h\ge {c-1\over24}$.
\item The spectrum contains a continuum of states with integer spins.
\item The density of states is not always positive. In particular, the degeneracy of the state with $h=\bar h={c-1\over24}$ is $-6$.
\end{itemize}
Since the MWK partition function has no non-vacuum state with twist below $c-1\over 16$, our argument in Section \ref{sec:positive} suggests that the density of states must turn negative in the regime \eqref{negativity}, in addition to the known negativity at $h=\bar h={c-1\over24}$. We will show that this is exactly the case.

Below we review the  density of states for the MWK partition function. 
Instead of using the spin $j$ and the twist $2t$, we will follow the convention in \cite{Keller:2014xba} and use the variables  $e$ and $j$ defined as
\ie
e= h+\bar h - {c-1\over 12}\,,~~~~~j=h-\bar h\,.
\fe
The  density of states for the MWK partition function receives contributions from the $PSL(2,\mathbb{Z})$ sums from each of the four terms ${1\over|\eta(q)|^2}q^{E_L}\bar q^{E_R}$ in \eqref{vacchi}
\ie\label{MWKrho}
\rho^{\rm MWK}_j(e)  = \rho^{ -{c-1\over12}  ,0} _j(e)
- \rho^{  - {c-13\over12}  ,1 }_j (e)- \rho^{  - {c-13\over12}  ,-1 }_j (e) 
+\rho^{ -{c-25\over12}  ,0} _j(e) \,,
\fe
where the superscripts $E,J$ of $\rho^{E,J}_j(e)$ are defined analogously as $E=E_L+E_R ,J  =E_L-E_R$ for each of the ``seed'' terms ${1\over |\eta(q)|^2}q^{E_L}\bar q^{E_R}$  in \eqref{vacchi}.

Each of the four terms in \eqref{MWKrho} is further written as an infinite sum:
\ie
\rho^{E,J}_{j} (e) = \sum_{m=0}^\infty \rho^{E,J}_{j,m}(e)\,.
\fe
When $j\neq0$, which is the case of interest, $\rho_{j,m}^{E,J}(e)$ is given by
\ie\label{rho}
\rho^{E,J}_{j,m}(e) = Z_{j,J}(m+1/2)\, {2^{3m+1} \pi^{2m} \over (2m)!} 
|j|^{m-1} \, \nu_{m}^{E,J}(e/|j|)\,,
\fe
where the  function $\nu^{E,J}_{m}(t)$ will be defined momentarily.\footnote{$\nu_m^{E,J}(e/|j|)$ is denoted as $\nu_{j,m}(e)$ in \cite{Keller:2014xba}.}  
$Z_{j,J}(m+1/2)$ is the Kloosterman zeta function defined by a Dirichlet series
\ie
Z_{j,J}(m+1/2)  =  \sum_{s=1}^\infty s^{-2(m+1/2)}  S(j,J;s)\,.
\fe
When $m=0$, the above series diverges and the Kloosterman zeta function $Z_{j,J}(1/2)$ is defined by analytic continuation and finite of order ${\cal O}(j^2 J^2)$ \cite{Maloney:2007ud,Keller:2014xba}.  For our purposes, the dominant contributions come from the higher $m$ terms, so we will not be concerned about the explicit regularized values of $Z_{j,J}(1/2)$. 

Below we define the function $\nu^{E,J}_{m}(t)$. Since  all of the non-vacuum states in the MWK partition function have $h,\bar h\ge {c-1\over24}$, i.e. $e\ge |j|$, we will define the function $\nu^{E,J}_{m}(t)$ only for $t\ge 1$.  
We first define 
\ie
f_m(t) =  {1\over \sqrt{t^2-1}  }  \cosh (m \cosh^{-1}(t) )\,.
\fe
Next, we define the operator $D_t$ as
\ie
D_t f_k  = -E f_k - Jf_{k-1} - \frac J2 (1-\frac mk) (f_{k+1} -f_{k-1})\,,
\fe
which can be realized by a matrix action.  With these preparations, $\nu^{E,J}_m(t)$ is given by
\ie
\nu^{E,J}_m(t) = D_t^m  f_m(t)\,.
\fe

\subsection{Double Limit}

Here we consider a special case of the double limit \eqref{negativity} where we first take $\bar h $ to $c-1\over24$, and then take $j$ to be odd and large.  The function $\nu^{E,J}_m(t)$ simplifies in the limit $t= e/|j|\to1$:
\ie\label{appnu}
\nu^{E,J}_m(t)  \simeq \sqrt{1\over 2(t-1)} (-E-J)^m 
 \,.
\fe
Hence in this limit, we have\footnote{Strictly speaking, the $m=0$ term requires analytic continuation to make sense, but in the following we will only be interested in the large $m$ terms which dominate in the limit \eqref{negativity}.  }
\ie
\rho_j^{E,J}(e)& \simeq \sum_{s=1}^\infty {2 \over s}S(j,J;s) \sum_{m=1}^\infty  s^{-2m} {2^{3m} \pi^{2m}\over (2m)!} |j|^{m-1} (-E-J)^m \sqrt{j\over 4(\bar h-{c-1\over24})}\,.
\fe
In the large spin $j$ limit, the sum is dominated by large $m$, and we can approximate the sum in $m$ by an exponential:
\ie
\rho_j^{E,J}(e) \simeq 
\sum_{s=1}^\infty {1\over 2 s \sqrt{\, j\,(\bar h-{c-1\over24}) }}S(j,J;s)  \exp\left( {4\pi\over s} \sqrt{ {-E-J\over2} \,j\, }\right)\,.
\fe

Adding up the four terms in \eqref{MWKrho}, we obtain the following expression for the MWK density of states in the limit where we take $\bar h\to {c-1\over24}$ first and then $j\to \infty$:
\begin{align}\label{MWKlimit}
\rho_j(e)& \simeq \sum_{s=1}^\infty  {1\over 2s \sqrt{j (\bar h-{c-1\over24})}  } 
\left  [    S(j,0;s)  \exp\left( {4\pi\over s} \sqrt{ {c-1\over24}  \, j \,} \right)
+ S(j,0;s)  \exp\left( {4\pi\over s} \sqrt{ {c-25\over24}  \, j \,} \right)\right.\nn\\
&\left.
- S(j,-1;s)  \exp\left( {4\pi\over s} \sqrt{ {c-1\over24}  \, j \,} \right)
- S(j,1;s)  \exp\left( {4\pi\over s} \sqrt{ {c-25\over24}  \, j \,} \right)
\right]
\end{align}
which is precisely our formula \eqref{eq:rho0} for the asymptotic density of states in this limit.\footnote{In comparing the above with \eqref{eq:rho0}, there is a relative factor of 2 coming from the Jacobian factor when we change variables from $t,j$ to $e,j$.} 
In particular, the $s=1$ term in \eqref{MWKlimit} vanishes and the density of states is dominated by the $s=2$ term, which is negative  in this limit  as argued in Section \ref{sec:positive}. Furthermore, we can show that the MWK density of states is negative in the more general regime \eqref{negativity} as predicted by \eqref{eq:rho0} by keeping the subleading term of $\nu^{E,J}_m(t)$ in the $t\to 1$ limit, namely
\begin{align}
\nu^{E,J}_m(t) &= \sqrt{\frac{1}{2(t-1)}}(-E-J)^m  \\&~~~+\frac{(2m-1)(-E-J)^{m-1}((2m+1)(-E)+(2m-1)J)\sqrt{t-1}}{4\sqrt2} + \mathcal{O}\((t-1)^{\frac32}\).\nn
\end{align}
This  gives the subleading term in $(\bar h - \frac{c-1}{24})$ when plugged into $\rho_j(e)$. In particular, this accounts for the leading nonzero contribution at $s=1$. We have also confirmed our prediction \eqref{eq:rho0} numerically with the MWK density of states. 

In \cite{Keller:2014xba}, the authors show that the density of states is positive if we fix  $e$ and $j$, and then take $c$ to be large.  Here we uncover the negative density of states in a different regime \eqref{negativity} where both $e$ and $j$ are taken to be much large than $c$.  
This new regime of  negative density of states makes it more challenging to correct the MWK partition function to a unitary, physical partition function.  

To cancel the negative density of states in the regime \eqref{negativity} of the MWK partition function without ruining modular invariance, one tentative candidate is to add $N$ copies of the $PSL(2,\mathbb{Z})$ sum of the state $(h,\bar h)= ({c-1\over32}, {c-1\over 32})$ above the vacuum.  
The ``seed'' term of this addition to the partition function is  ${N\over |\eta(q)|^2}q^{ {c-1\over 32} - {c-1\over 24}} \bar q^{ {c-1\over 32} - {c-1\over 24}}  ={N\over |\eta(q)|^2} q^{- { c-1\over 96}} \bar q^{\,- {c-1\over 96}}$, and therefore contributes to the density of states by $N\rho^{ E = - {c-1\over 48} ,J=0}$. 
As shown in \cite{Keller:2014xba}, $\rho_j^{E,0}(e)$ (with $E<0$) is positive everywhere  except for a negative delta function at $e=0$ with $j=0$.   This delta function negativity can be canceled by adding, for example, the modular invariant partition function  of the $c=1$ self-dual boson (see Section 4.2 of \cite{Keller:2014xba}). 
Putting everything together, let us consider the final density of states
\ie
\rho^{\rm MWK}_j(e)  + N  \rho^{ - {c-1\over 48} , 0}(e)  +  (N+6) \rho^{c=1}_j(e)
\label{eq:mwkplusstuff}
\fe
where $\rho^{c=1}_j(e)$ is the density of states for the $c=1$ self-dual boson, whose vacuum is now interpreted as a $h=\bar h = {c-1\over 24}$ state. 
The term $6 \rho^{c=1}_j(e)$  is there to cancel the negative delta function density of states at $h=\bar h = {c-1\over24}$ coming the MWK partition function itself. 
For $N>1$,  the above density of states  appears to be positive  at large $c$ in all regimes we have considered, and gives a  modular invariant  partition function.  
It has a vacuum, a finite number of states at $h=\bar h = {c-1\over 32}$, and a continuum starting at $h,\bar h \ge {c-1\over 24}$. 
The twist gap $2t_{\rm gap}$ and the gap in the scaling dimension $\Delta_{\rm gap}$ are both ${c-1\over 16}$.  
It would be interesting to prove that this density of states is indeed positive everywhere.\footnote{Note that there is a simple alternative to (\ref{eq:mwkplusstuff}) where instead of adding scalars of twist $\frac{c-1}{16}$ to the MWK partition function, we add twist $\frac{c-1}{16}$ states with arbitrary spin (plus $PSL(2,\mathbb Z)$ images). If the spectra of these partition functions are also positive-definite, they would have twist gap $\frac{c-1}{16}$, and scaling gap no greater than $\frac{c-1}{12}$.}

\section{Supersymmetric Generalization}\label{sec:N=1}

Our arguments in this paper can be generalized to 2d CFTs with any chiral algebra. 
In this section we will perform a similar analysis for the $\mathcal{N}=1$ super-Virasoro algebra.
Recall that there are four partition functions we can consider, depending on the four spin structures on the torus, which correspond to (anti)periodic boundary conditions of the fermions in the space and time directions.\footnote{Here we assume for simplicity that left and right moving fermions have identical spin structures. It would be interesting to explore modular constraints by considering partition functions with mixed spin structures.} These correspond to partition functions restricted to the Neveu-Schwarz or Ramond sectors; and with or without a $(-1)^F$ insertion. Three of these partition functions are related by $PSL(2,\mathbb Z)$ transformations, and the remaining one is the Witten index.

In this section we will focus on the partition function with anti-periodic boundary conditions for the fermions in both directions on the torus, namely
\be
Z(q, \bar q) = \text{Tr}_{\mathcal{H}_{\text{NS,NS}}}\(q^{L_0-\frac{c}{24}} \bar q^{\overline{L}_{0} - \frac{c}{24}}\).
\ee
This function is invariant under a subgroup of $PSL(2,\mathbb Z)$, generated by the group elements $S$ and $T^2$. We will denote this group as $\Gamma_{\theta}$, which can also be defined by $SL(2,\mathbb Z)$  matrices $\begin{pmatrix} a & b \\ c & d\end{pmatrix}$ with $a+d$ and $b+c$ both even. The characters for the $\mathcal{N}=1$ super-Virasoro algebra with $c>\frac32$ under this spin structure are
\begin{align}
\chi_0^{\mathcal{N}=1}(q) &= q^{-\frac{c-\frac32}{24}} (1-q^\frac12) \frac{\eta(q)}{\eta(q^2)\eta\( q^{1/2}\)} ~~~ \text{vacuum}\nn\\
\chi_h^{\mathcal{N}=1}(q) &= q^{h-\frac{c-\frac32}{24}}  \frac{\eta(q)}{\eta(q^2)\eta\( q^{1/2}\)} ~~~~~~~~~~~~~h>0.
\end{align}
We can similarly define a modular kernel $K_S^{\mathcal{N}=1}(h)$ as
\be
\chi_0^{\mathcal{N}=1}(q') = \int_0^{\infty} dh K_S^{\mathcal{N}=1}(h) \chi_h^{\mathcal{N}=1}(q)
\ee
where $q'$ is the $S$ transform of $q$. It is given by
\begin{align}
K_S^{\mathcal N=1}&(h)= \sqrt{\frac{2}{h - \frac{c-\frac32}{24}}}\theta\(h-\frac{c-\frac32}{24}\)
\\&\times\left[ \cosh\(4\pi\sqrt{\(\frac{c-\frac32}{24}\)\( h - \frac{c-\frac32}{24}\)}\) - \cosh\(4\pi\sqrt{\(\frac{c-\frac{27}2}{24}\)\( h - \frac{c-\frac32}{24}\)}\)\right]. \nn
\label{eq:Khsusy}
\end{align}

By the same argument in Section \ref{sec:CardyPlus}, any 2d CFT with $\mathcal{N}=1$ super-Virasoro symmetry and a nonzero twist gap under the $\mathcal{N}=1$ super-Virasoro algebra obeys a Cardy-like formula with extended validity. In particular the density of super-Virasoro primaries is given by
\begin{equation}
\rho^0_{j,1}(\bar h) = K_S^{\mathcal{N}=1}(\bar h +j) K_S^{\mathcal{N}=1}(\bar h),
\label{eq:susyExtCardy}
\end{equation}
which is valid whenever $\bar h + j \gg c$, $\bar h > \frac{c-\frac32}{24}$.

However, just as for the non-supersymmetric case, (\ref{eq:susyExtCardy}) is not all we can learn from the lightcone modular bootstrap. Let us consider a more general setup where we set $\tau$ and $\bar\tau$ as in \eqref{ttb}, 
in the limit where $\beta$ goes to zero with $\overline \beta$ fixed. Again, $r$ and $s$  in  \eqref{ttb} are coprime integers, and $s \geq 1$; we will now in addition demand that $r$ and $s$ have different parity. We can consider a $\Gamma_{\theta}$ transformation 
\be
\begin{pmatrix} a(r,s) & b(r,s) \\ s & -r \end{pmatrix}.
\ee
For consistency, $a(r,s)$ must satisfy the following properties: If $s$ is even, $a(r,s) = -\(r^{-1}\)_{2s}$. If $s$ is odd, $a(r,s) = -\(r^{-1}\)_s$ and chosen so that $a(r,s)$ is even. We then obtain the following crossing equation (similar to \eqref{eq:finalcrossing} for the Virasoro case):
\be
\label{eq:n1finalCrossing}
&\sum_{\substack{j \geq 0 \\ j\in \mathbb{Z}/2}} \exp\left[-\(\beta - \frac{2\pi i r}{s}\)j\right] \int d\bar h \rho_j(\bar h) \exp\left[-\overline\beta\(\bar h - \frac{c-\frac32}{24}\)\right]\\
&=\frac{2\pi}{s\sqrt{\beta\overline\beta}} e^{\frac{4\pi^2(c-\frac32)}{24s^2\beta}} e^{\frac{4\pi^2(c-\frac32)}{24s^2\overline\beta}} (1-e^{-\frac{2\pi^2}{s^2\beta} + \frac{\pi i a(r,s)}{s}})(1-e^{-\frac{2\pi^2}{s^2\overline\beta} - \frac{\pi i a(r,s)}{s}}) + \mathcal{O}\(e^{\frac{4\pi^2}{s^2\beta}\(\frac{c-\frac32}{24}-t_{\text{gap}}\)}\).\nn
\ee
In (\ref{eq:n1finalCrossing}), the sum over $j$ runs over both integers and half-integers. 

We will find a continuous solution to (\ref{eq:n1finalCrossing}), in a similar manner as in Section \ref{sec:solC}. We define a $\Gamma_{\theta}$ Kloosterman sum as 
\ie\label{kloostermanGammaTheta}
S^{\Gamma_\theta}(j, J ; s)  = \sum_{\substack{r:\,\text{gcd}(r,s)=1  \\ 0 \leq r<2s \\ r+s~\text{odd}} }  \exp\left(\pi i  \, {  2rj - a(r,s) J\over s}\right)\,,
\fe
which satisfies 
\be
\sum_{\substack{j=0\\j \in \mathbb{Z}/2}}^{\text{lcm}(s,s')-\frac12} S^{\Gamma_\theta}(j,J,s') e^{-\frac{2\pi i r j}{s}} = 2s e^{-\frac{\pi i a(r,s) J}s} \delta_{s,s'} .
\ee
We also define the functions
\begin{align}
d_0^{\mathcal{N}=1}(h, s) &= \sqrt{\frac{2}{s(h-\frac{c-\frac32}{24})}}\cosh\(\frac{4\pi}{s}\sqrt{\(h-\frac{c-\frac32}{24}\)\(\frac{c-\frac32}{24}\)}\)\theta(h-\frac{c-\frac32}{24}) \nn\\
d_1^{\mathcal{N}=1}(h, s) &= \sqrt{\frac{2}{s(h-\frac{c-\frac32}{24})}}\cosh\(\frac{4\pi}{s}\sqrt{\(h-\frac{c-\frac32}{24}\)\(\frac{c-\frac{27}2}{24}\)}\)\theta(h-\frac{c-\frac32}{24})\,,
\end{align}
and 
\begin{align}
\rho_{j,s}^{\mathcal{N}=1}(h, \bar h) &= S^{\Gamma_\theta}(j,0;s)  d_0^{\mathcal{N}=1}(h, s) d_0^{\mathcal{N}=1}(\bar h,s) - S^{\Gamma_\theta}(j,-1;s) d_0^{\mathcal{N}=1}(h,s) d_1^{\mathcal{N}=1}(\bar h,s)\nn\\
&~~ - S^{\Gamma_\theta}(j,1;s)d_1^{\mathcal{N}=1}(h,s) d_0^{\mathcal{N}=1}(\bar h,s) +  S^{\Gamma_\theta}(j,0;s)d_1^{\mathcal{N}=1}(h,s) d_1^{\mathcal{N}=1}(\bar h,s)\,.
\end{align}
A solution to (\ref{eq:n1finalCrossing}) at large spin is then given by
\be
\rho_j^0(\bar h) = \sum_{s'=1}\rho_{j,s'}^{\mathcal{N}=1}(\bar h +j , \bar h)\,,~~~~j\gg c\,.
\label{eq:rho0SUSY}
\ee
The solution (\ref{eq:rho0SUSY}) again has the interesting property that for large odd (or half-integer) spin, with $\bar h$ exponentially close to $\frac{c-\frac32}{24}$, there is a negative degeneracy of states. This can be cured with a single operator whose twist is at or below $\frac{c-\frac32}{16}$.

\section*{Acknowledgements}
We thank S. Collier, T. Hartman, S. Kachru, Z. Komargodski, Y.-H. Lin, R. Mahajan, A. Maloney, H. Maxfield, D. Mazac, B. Mukhametzhanov, E. Perlmutter, L. Rastelli, D. Simmons-Duffin, D. Stanford, H. Verlinde, and E. Witten for interesting discussions.  
We thank T. Hartman, C. Keller, A. Maloney, H. Maxfield, and X. Yin  for comments on a draft. 
The work of N.B. is supported in part by the Simons Foundation Grant No. 488653. 
The work of H.O. is supported in part by
U.S.\ Department of Energy grant DE-SC0011632,
by the World Premier International Research Center Initiative,
MEXT, Japan,
by JSPS Grant-in-Aid for Scientific Research C-26400240,
and by JSPS Grant-in-Aid for Scientific Research on Innovative Areas
15H05895.
The work of  S.H.S. is supported  by the National Science Foundation grant PHY-1606531 and by the Roger Dashen Membership. The work of Y.W.
is supported in part by the US NSF under Grant No. PHY-1620059 and by the Simons
Foundation Grant No. 488653.  
N.B. and H.O. thank the Aspen Center for Theoretical Physics, which is supported by
the National Science Foundation grant PHY-1607611,  where part of this work was done.

\appendix

\section{Sum Estimation}
\label{sec:sumest}

Consider a complex function $f(z)$  that satisfies the following properties:
\begin{itemize}
	\item $f(z)$ is analytic on the half-plane ${\rm Re}\,z\geq 0$.
	\item $ \lim_{|{\rm Im}\,  z|\to \infty}e^{-2\pi|{\rm Im}\,z|}|f(z)|= 0$  uniformly in any finite interval $(\D,\Lambda)$ of ${\rm Re}\,z$.
\end{itemize} 
From a contour argument, one can derive the Abel-Plana formula (see section 3 in \cite{olver} or section 13.14 in \cite{hardy}) which relates the discrete sum to the integral of $f(z)$,
\ie
&\sum_{k=0}^\infty f(k)-\int_0^\infty dz f(z)
\\=&{1\over 2} f(0)+  i\int_{0}^\infty dz  {f(iz)-f(-iz)\over e^{2\pi z}-1}
+\lim_{\Lambda   \to +\infty}\left(
{1\over 2} f(\Lambda)+  i\int_{0}^\infty dz  {f(\Lambda-iz)-f(\Lambda+iz)\over e^{2\pi z}-1}
\right)
.
\label{APF}
\fe

One application of the formula arises in the main text where we replace the sum over spin $j$ by an integral over $j$ in \eqref{checkextcardy}. In this case, we choose 
\ie
f(j)=e^{-\B j} \int_{c-1\over 24}^\infty d\bar h \frac{e^{4\pi \sqrt{(\bar h+j)(\frac{c-1}{24})}}K_S(\bar h)}{\sqrt{\bar h + j - \frac{c-1}{24}}} \exp\left[ -\overline\beta (\bar h-{c-1\over24}) \right]
\fe
which clearly satisfies the criterion for the Abel-Plana formula and in addition $\lim_{\Lambda\to +\infty}f(\Lambda+iy)=0$. Moreover the RHS of \eqref{APF} is finite in the limit $\B \to 0$, thus we can freely replace the sum with the integral (or vice versa) without affecting the exponentially dominating terms of the form ${\cal O}(e^{a \over \B})$ for some $\B$-independent constant $a$.

\section{Crossing Solution}
\label{sec:plugin}

In this appendix we show that (\ref{eq:rho0}) solves the crossing equation (\ref{eq:finalcrossing}).  In particular, we show that for the modular crossing equation (\ref{eq:finalcrossing}) labeled by $s$ (and for all coprime $1\le r<s$), only the $\rho_{j, s'=s}$ term in (\ref{eq:rho0}) contributes to  the leading term on the RHS of (\ref{eq:finalcrossing}) in the $\beta\to0$ limit.  

The LHS of (\ref{eq:finalcrossing}) is given by  (see end of Section~\ref{sec:solC} for the range of $s'$)
\begin{align}
&\sum_{j=0}^{\infty} e^{\frac{2\pi i r j}s} \exp\left[{-j\beta}\right] \int_{t_{\text{gap}}}^{\infty}  d\bar h \rho_{j}(\bar h) \exp\left[{-\overline\beta(\bar h-\frac{c-1}{24})}\right]  \nn\\
&=\sum_{s'}\sum_{j=0}^{\infty} e^{\frac{2\pi i r j}s} \exp\left[{-j\beta}\right]\int_{t_{\text{gap}}}^{\infty}  d\bar h \exp\left[{-\overline \beta(\bar h-\frac{c-1}{24})}\right]  \nn\\ &~~~~~~~\times \left[\, 
 S(j,0;s')d_0(\bar h+j,s')d_0(\bar h,s') - S(j,-1;s' ) d_0(\bar h +j,s')d_1(\bar h,s') \right.\nn\\
&\left.~~~~~~~~~~~~~~~~ - S(j,1;s') d_1(\bar h+j,s')d_0(\bar h,s')+ S(j,0;s') d_1(\bar h+j,s')d_1(\bar h,s')
 \, \right].
\label{eq:mess}
\end{align}
Let us consider the first term coming from the square bracket on the RHS of (\ref{eq:mess}). We rewrite $j=\tilde j + k~ \text{lcm}(s, s')$ where $k\geq 0$ and $0\leq \tilde j \leq \text{lcm}(s,s')-1$ to get:
\begin{align}
&\sum_{s'} \sum_{j=0}^{\infty} e^{\frac{2\pi i r j}s} \exp\left[{-j\beta}\right] \int_{t_{\text{gap}}}^{\infty}  d\bar h
S(j,0;s') d_0(\bar h+j,s')d_0(\bar h,s') \exp\left[-{\overline\beta(\bar h-\frac{c-1}{24})}\right]  \nn\\
&= \sum_{s'} \sum_{\tilde j=0}^{\text{lcm}(s,s')-1} e^{\frac{2\pi i r \tilde j}s} \,S(\tilde j , 0 ;s')
\sum_{k=0}^{\infty}  \exp\left[{-(\tilde j + k~\text{lcm}(s,s'))\beta}\right] \nn\\ &~~~~~~~~~\times \int_{t_{\text{gap}}}^{\infty}d\bar h \, d_0(\bar h+\tilde j + k~\text{lcm}(s,s'),s') d_0(\bar h,s') \exp\left[{-\overline\beta(\bar h-\frac{c-1}{24})}\right].
\label{eq:mess2}
\end{align}
If we replaced the sum over $k$ with an integral in (\ref{eq:mess2}), we could shift variables $k\rightarrow k -\frac{\tilde j}{\text{lcm}(s,s')}$. In Appendix \ref{sec:sumest}, we show that the correction coming from changing the sum to an integral approaches a constant as $\beta$ goes to $0$. Since we only aim to reproduce the divergence as $\beta\to0$ on the RHS of \eqref{eq:finalcrossing}, we can approximate the sum in $k$ by an integral:
\begin{align}
\sum_{s'} 
\left[ 
\sum_{\tilde j=0}^{\text{lcm}(s,s')-1} e^{\frac{2\pi i r \tilde j}s}S(\tilde j , 0 ;s')
\right]
 \int_0^{\infty} dk \int_{t_{\text{gap}}}^{\infty} &d\bar h  \,  
 d_0(\bar h+k~\text{lcm}(s,s'),s') d_0(\bar h,s') \label{eq:mess3}\\
  &\times\exp\left[{-\beta k~\text{lcm}(s,s')}\right]\exp\left[{-\overline\beta(\bar h-\frac{c-1}{24})}\right] \notag
\end{align}

The sum over $\tilde j$ in (\ref{eq:mess3}) can now be done using (\ref{eq:abprop}), which is proportional to $\delta_{s,s'}$. 
It follows that only the $\rho_{j, s'=s}$ term in the sum \eqref{eq:rho0} contributes to the modular crossing equation \eqref{eq:finalcrossing} labeled by $s$.  
The $s'=s$ term then gives:
\begin{align}
&s \int_0^{\infty} dk \int_{t_{\text{gap}}}^\infty d\bar h d_0(\bar h+ks,s) d_0(\bar h,s)\exp\left[{-\overline\beta(\bar h-\frac{c-1}{24}) - \beta k s }\right] \nn\\
&=  \int_{t_{\text{gap}}}^\infty d\bar h \exp\left[\beta \bar h \right] \int_{\bar h}^{\infty} dk d_0(k,s) d_0(\bar h,s)\exp\left[{-\overline\beta(\bar h-\frac{c-1}{24}) - \beta k}\right] \nn\\
&= \sqrt{\frac{2\pi}{s\beta}} e^{\frac{4\pi^2(c-1)}{24s^2\beta}} \int_{t_{\text{gap}}}^\infty d\bar h \exp\left[(\beta-\overline\beta) (\bar h-\frac{c-1}{24}) \right] d_0(\bar h,s) \nn\\
&\simeq \sqrt{\frac{2\pi}{s\beta}} e^{\frac{4\pi^2(c-1)}{24s^2\beta}} \int_{t_{\text{gap}}}^\infty d\bar h \exp\left[-\overline \beta (\bar h-\frac{c-1}{24}) \right] d_0(\bar h,s) \nn\\
&= \frac{2\pi}{s\sqrt{\beta\overline\beta}} e^{\frac{4\pi^2(c-1)}{24s^2\beta}}e^{\frac{4\pi^2(c-1)}{24s^2\overline\beta}}
\label{eq:mess4}
\end{align}
where we use $\simeq$ to mean equal in the limit $\beta \rightarrow 0$.  We have succeeded in reproducing one of the divergent terms on the RHS of \eqref{eq:finalcrossing}.

We can evaluate the remaining three terms in (\ref{eq:mess}) following the same steps. When combined, we finally get
\be
\frac{2\pi}{s\sqrt{\beta\overline\beta}} e^{\frac{4\pi^2(c-1)}{24s^2\beta}}e^{\frac{4\pi^2(c-1)}{24s^2\overline\beta}}(1-e^{-\frac{4\pi^2}{s^2\beta}-\frac{2\pi i \(r^{-1}\)_s} s})(1-e^{-\frac{4\pi^2}{s^2\overline\beta}+\frac{2\pi i\(r^{-1}\)_s} s})
\ee
precisely matching (\ref{eq:finalcrossing}).

\section{Accumulation Point in Twist}
\label{sec:accumulation}
In Section \ref{sec:positive}, we gave a suggestive argument that the twist gap $2t_{\text{gap}}$ cannot exceed $\frac{c-1}{16}$. However, there we assumed that the contribution of the lowest twist non-vacuum operator to the  density of large spin states came from a finite number of low-twist operators. It is possible that the twist gap comes from an accumulation point in twist. In that case, it is not obvious that the contribution to the high-spin density of states has the $\bar h$-dependence in \eqref{lowesttwistguy}.

We start with the modular constraint \eqref{SConstraint} rewritten as
\ie
&\sum_{j=0 }^\infty e^{ -\beta j } \int_{0}^\infty d\bar h \, \rho_j (\bar h) \exp\left[ -\overline\beta (\bar h-{c-1\over24}) \right]\\
=&  
\sum_{j=0 }^\infty e^{ -\beta j }
\int_0^\infty d\bar h  \rho_j^0 (\bar h)   \exp \left[{-\overline\B(\bar h -{c-1\over 24})}\right]
+{\cal O}(e^{{4\pi^2 \over \beta}({c-1\over24} - t_{\rm gap})  })\,.\label{modeqn}
\fe

Recall that $\rho_j(\bar h)$ is the physical density of states \eqref{phyden} and $\rho_j^0(\bar h)$ is defined in \eqref{eq:rho0}. 
We perform an inverse Laplace transform in $\B$,\footnote{Here and below we assume that the inverse Laplace transform of the error term in \eqref{modeqn} with respect to $\B$ is dominated by the inverse Laplace transform of $e^{{4\pi^2 \over \beta}({c-1\over24} - t_{\rm gap})  }$  for large $j$. To prove this requires a refinement of the argument in \cite{Mukhametzhanov:2019pzy} with a finite twist gap (or irrationality) condition imposed. }
\ie
& \int_{0}^\infty d\bar h \, \rho_j (\bar h) \exp\left[ -\overline\beta (\bar h-{c-1\over24}) \right]=  
\int_{0}^\infty d\bar h  \, \rho_j^0 (\bar h)   \exp \left[{-\overline\B(\bar h -{c-1\over 24})}\right]
+{\cal O}({e^{{4\pi}\sqrt{j({c-1\over24} - t_{\rm gap})  }} }).
\label{ILV1}
\fe
It is then tempting to equate $\rho_j(\bar h)$ with $\rho_j^0(\bar h)$, but  the inverse Laplace transform with respect to $\overline{\B}$ is only unique up to a measure zero set of $\bR^+$. 
Nonetheless, two piecewise continuous functions with the same Laplace transforms agree on the subset of $\bR^+$ where they are both continuous (see, for example, \cite{doetsch}).
To utilize this uniqueness property, we consider the integrated spectral density for $\rho_j(\bar h)$ defined as
\ie
F_j(\bar h)=\int_0^{\bar h}   d\bar h' \rho_j(\bar h')
\fe
which is piecewise continuous. 
Similarly we define  $F_j^0(\bar h)$ for $\rho^0_j(\bar h)$. By performing integration by parts on \eqref{ILV1}, we obtain
\ie
& \int_{0}^\infty d\bar h \, F_j (\bar h) \exp\left[ -\overline\beta (\bar h-{c-1\over24}) \right]\\
=&  
\int_0^\infty d\bar h \, F_j^0 (\bar h)    \exp \left[{-\overline\B(\bar h -{c-1\over 24})}\right]
+{\cal O}({e^{{4\pi}\sqrt{j({c-1\over24} - t_{\rm gap})  }} }),
\label{ILV2}
\fe
where we used $\lim_{\bar h\to \infty}F_j(\bar h) e^{-\overline \B \bar h}=0$ as evident from the usual Cardy formula. Next performing the inverse Laplace transform in $\overline \B$, we obtain 
\ie\label{Fj}
F_j (\bar h)  
=
F_j^0 (\bar h)    
+{\cal O}\left({e^{{4\pi}\sqrt{j({c-1\over24} - t_{\rm gap}) }}  }\right)
\fe
where the equality holds away from the discontinuities of the LHS. 
We define the function
\ie
g_j (\bar h )  \equiv F_j (\bar h ) - F_j^0(\bar h)\,.
\fe
It satisfies the following two properties:
\begin{enumerate}
\item  $\lim_{j\to\infty} { g_j (\bar h )  \over \exp \left[4\pi \sqrt{ j ({c-1\over 24} -t_{\rm gap})}\right]} = G(\bar h)$ where $G(\bar h)$ is some finite function of $\bar h$.
\item  $g_j(\bar h)$ is finite at every $\bar h$, and in particular $\lim_{\bar h \to {c-1\over24} }  g_ j(\bar h) $ is finite. 
\end{enumerate}
The first property follows from \eqref{Fj} while the second follows from the finiteness of the number of states up to twist $\bar h$ at a given spin $j$.

The question we are interested in is the large $j$ growth of $g_j(\bar h)$ in the double limit \eqref{negativity}. To make it more precise, let us choose any function $f(j)$ such that
\ie
0< f( j )  -{c-1\over 24}  < {1\over 8\pi^2}  \exp\left[ -2\pi \sqrt{  {c-1\over24}\, j\, }\right],~~j:~\text{odd}\,.
\fe
We would like to know the large $j$ behavior of  $g_ j (\bar h = f( j) )$.  
In this limit, we have shown in Section \ref{sec:positive} that 
\ie
F_j^0 (\bar h)     \sim - e^{\pi \sqrt{j {c-1\over 24}}}
\fe
If we naively commute the two limits $j\to\infty$ and $\bar h\to{c-1\over24}$, we would have claimed that $g_j(\bar h=f(j))$ grows no faster than $e^{4\pi \sqrt{ j ({c-1\over24}  -t_{\rm gap}})}$, i.e. 
\ie
\label{doubledlimit}
\lim_{j\to \infty}  {g_ j (f(j))  \over \exp\left[ 4\pi \sqrt{j ({c-1\over24} - t_{\rm gap})  } \right] }   
\fe
is finite. If this were the case, 
then for the negativity of $F_j^0(\bar h) \sim -e^{ \pi \sqrt{{c-1\over24} j }}$ to be  canceled by the growth $e^{4\pi \sqrt{ j ({c-1\over24}  -t_{\rm gap}})}$, the twist has to be no bigger than
\ie
t_{\rm gap}\leq \(\frac{15}{16}\)\({c-1\over 24} \).
\fe
If it can be proven that  \eqref{doubledlimit} is not only  finite, but vanished as $\sqrt{\bar h -\frac{c-1}{24}}$ (as would be the case if there were a finite number of low-twist operators), we would recover the original claim of $t_{\text{gap}} \leq \frac{c-1}{32}$ in Section \ref{sec:positive}. 
However, \eqref{doubledlimit} involves a double limit on both $\bar h$ and $j$, whose behavior does not follow from the two properties of $g_j(\bar h)$ above. 
We leave a rigorous derivation  of the large $j$ behavior of $g_j(f(j))$ for the future.

\section{Modular Transformation of Virasoro Characters}
\label{sec:modularkernel}
In this appendix, we derive simplified formulae for the  modular kernel of general $PSL(2,\mathbb Z)$ transformations for Virasoro characters at  central charge $c>1$. 
See \cite{Jego:2006ta,Ribault:2016sla} for a nice review on this subject. We will mostly follow the conventions of \cite{Jego:2006ta} here and focus on the holomorphic side.

It is convenient to  parametrize the CFT central charge as, 
\ie
c=1+6Q^2,\quad 
Q=b+{1\over b}
\fe
where $0< b< 1$ for $c>25$ or $b$ is a complex phase with $|b|=1$ for $1<c\leq 25$, 
inspired by the Liouville theory. Similarly, we label Virasoro primaries of weight $h$ by the
Liouville momenta
\ie
\A\equiv {Q+i p\over 2},\quad h={\A (Q-\A)}
\fe
such that for $p\in \mathbb{R}^+$, we have a non-degenerate primary  with character and weight
\ie
\chi_p(\tau) ={q^{p^2\over 4}\over \eta(\tau)},\quad h_p={Q^2+p^2\over 4}.
\fe
For imaginary $p=i(m/b+nb)$ with $m,n\in \mathbb N$, we have a degenerate primary (which has a single null vector at level $mn$ for generic $b$) with character and weight   
\ie
\chi_{m,n} ={q^{-{(m/b-nb)^2\over 4}}-q^{-{(m/b+nb)^2\over 4}}\over \eta(\tau)},\quad h_{m,n}={Q^2-(m/b+nb)^2\over 4} .
\fe 
In particular, the  vacuum is identified with the $m=n=1$ degenerate primary
\ie
\chi_{\rm vac}\equiv \chi_{1,1}.
\fe
The $S$-transformation of the Virasoro characters are particularly simple in Liouville notation,
\ie
\chi_p(-1/\tau)&=\int_0^\infty dp'\, S_{p}{}^{p'}\chi_{p'},\quad S_{p}{}^{p'}=\sqrt{2}\cos(\pi pp')
\\
\chi_{m,n}(-1/\tau)&=\int_0^\infty dp'\, S_{m,n}{}^{p'}\chi_{p'},\quad S_{m,n}{}^{p'}=2\sqrt{2}\sinh (\pi m p'/b)\sinh (\pi n bp')
\label{Sk}
\fe
while the $T$-transformation   gives a phase
\ie
\chi_p(\tau+1)=e^{{\pi i\over 2}p^2}e^{\pi i \over 12}\chi_p(\tau),\quad 
\chi_{m,n}(\tau+1)=e^{-{\pi i\over 2}(m/b+nb)^2}e^{\pi i \over 12}\chi_{m,n}(\tau).
\label{Tk}
\fe

For a given element $\C \in PSL(2,\mathbb Z)$ generated by a sequence of $S$ and $T$ transformations, it is straightforward to compose the integrals and derive the modular kernel ${\bf K}_\C$.\footnote{Note that the modular crossing kernel ${\bf K}_\C$ here differs from $K_\C$ in the main text (e.g. \eqref{KS}) by the choice of integration measure over Virasoro modules. For example ${ K}_\C(h_p)={1\over\sqrt{h_{p}-{c-1\over 24}}} {\bf K}_\C({\rm vac};p)$ for non-degenerate modules labeled by $p$.} In particular
\ie
{\bf K}_S({m,n};p)=S_{m,n}{}^{p},\quad {\bf K}_S(p;p')=S_{p}{}^{p'}.
\fe
For illustration, let us look at the $ST^nS$ transformation (with $n \in {\mathbb N}$)
\ie
\C \cdot \tau ={\tau \over 1-n\tau}
\fe
of the identity character 
\ie
\chi_{\rm vac}(\C\cdot \tau)=2e^{-n\pi i \over 12}\int_{\mathbb{R}^+} dp'\sinh (\pi   p'/b)\sinh (\pi   bp') e^{n\pi ip'^2 \over 2}\int_\mathbb{R} dp e^{i\pi p p'} \chi_p(\tau).
\fe
We would like to perform the $p'$ integral first but as it stands above, the integral diverges exponentially. We can get around this by shifting the contour of the $p$-integral in the imaginary direction by $\Delta=1/b+b+\epsilon$ for $\epsilon>0$ \cite{Jego:2006ta}. Since there are no poles in $p$, we obtain
\ie
\chi_{\rm vac}(\C\cdot \tau)=2e^{-n\pi i \over 12}\int_{\mathbb{R}^+} dp'\sinh (\pi   p'/b)\sinh (\pi   bp') e^{n\pi i p'^2 \over 2}\int_{\mathbb{R}+i \Delta} dp e^{i\pi p p'} \chi_p(\tau)
\label{shiftedint}
\fe
and can now integrate over $p'$ using the following identities that involve the (complement) error function ${\rm erfc}(z)$ or equivalently the Faddeeva function $w(iz)$ 
\ie
{}&\int_{\mathbb{R}^+} dp' e^{\pi p'(\epsilon_1 /b+ \epsilon_2 b) }e^{n\pi i p'^2/2}  e^{i\pi p p'} 
={e^{ \pi i\over 4}\over  \sqrt{2n}} e^{z(p,\epsilon_1,\epsilon_2)^2} {\rm erfc}(z(p,\epsilon_1,\epsilon_2))
={e^{ \pi i\over 4}\over  \sqrt{2n}}  w(iz(p,\epsilon_1,\epsilon_2))
\fe
where
\ie
{\rm erfc}(z)\equiv {2\over \sqrt{\pi}}\int_z^\infty e^{-t^2} dt=1-{\rm erf}(z)
,\quad
w(z)=e^{-z^2}{\rm erfc}(-iz)
\fe
with
\ie
z(p,\epsilon_1,\epsilon_2)=e^{-3\pi i/ 4} \sqrt{\pi\over2 n} {(\epsilon_1 /b+ \epsilon_2 b+i p) }.
\label{zofp}
\fe
Now using the fact that $w(iz)$ is an entire function\footnote{For more properties of the Faddeeva function and related functions see Chapter 7 of \cite{NIST:DLMF}.} and the exponential suppression from $\chi_p(\tau)$ in the $p$-integral, we can deform the $p$-integration contour back to $\bR$ and \eqref{shiftedint} becomes
\begin{align}
\chi_{\rm vac}(\C\cdot \tau)
=&
{e^{ (3-n)\pi i\over 12}\over  2\sqrt{2n}}\sum_{\epsilon_{1,2}=\pm} \int_{\mathbb{R}} dp\,  w(iz(p,\epsilon_1,\epsilon_2)) \chi_p(\tau)\nn
\\
=&
{e^{ (3-n)\pi i\over 12}\over   \sqrt{2n}} \int_{\mathbb{R}} dp\,  (e^{z(p,1,1)^2}-e^{z(p,1,-1)^2}) \chi_p(\tau)\label{mtfe}
\\
=&
{e^{ (3-n)\pi i\over 12}\over   \sqrt{2n}}  \int_{\mathbb{R}^+} dp\,  
(e^{{\pi i\over 2n}(b+1/b+is)^2}-e^{{\pi i\over 2n}(b-1/b+is)^2} 
+
e^{{\pi i\over 2n}(b+1/b-is)^2}-e^{{\pi i\over 2n}(b-1/b-is)^2} 
)
\chi_p(\tau).\nn
\end{align}
In the second equality above,  we have used $w(-z)=2e^{-z^2}-w(z)$. 
We read off  the modular kernel ${\bf K}_{ST^n S}({\rm vac};p)$ from the last line of \eqref{mtfe} to be 
\ie
{\bf K}_{ST^n S}({\rm vac};p)  
=&{2e^{ (3-n)\pi i\over 12}\over   \sqrt{2n}} e^{\frac{2\pi i (c-1)}{24n}}  e^{- \pi i p^2 \over2 n}
\( \cosh{\(\pi p (b+1/b)\over n\)}
-
e^{-2\pi i \over  n}  \cosh{\(\pi p (b-1/b)\over n\)}
\)
\label{STnSk}
\fe

As a check of \eqref{STnSk}, note that in $PSL(2,\mathbb{Z})$, we have the identity $STS=T^{-1} S T^{-1}$. One can immediately write down the modular transformation of $\chi_{\rm vac}$ under $T^{-1} S T^{-1}$ using \eqref{Sk} and \eqref{Tk}:
\ie
\chi_{\rm vac}(\C\cdot \tau)=2\sqrt{2}e^{2\pi i \over 12}   e^{\frac{2\pi i (c-1)}{24} } \int_{\mathbb{R}^+} dp\sinh (\pi  p/b)\sinh (\pi   bp) e^{-\pi i p^2\over 2}  \chi_p(\tau)
\fe
and we see that this indeed matches \eqref{STnSk} when $n=1$.

Similarly, it is straightforward to derive the $ST^n S$ transformation of a non-degenerate Virasoro character labeled by $s\in \bR^+$,
\ie
\chi_s(\C\cdot \tau)=&{1\over 2}e^{-n\pi i \over 12}\int_{\mathbb{R}} dp' e^{i\pi s p'}  e^{n\pi ip'^2 \over 2}\int_{\mathbb{R}} dp e^{i\pi p p'} \chi_p(\tau)
\\
=&{1\over \sqrt{2n}}e^{\pi i (3-n) \over 12} \int dp e^{-{\pi i (p+s)^2\over 2n}}\chi_p(\tau)
\fe
thus
\ie
{\bf K}_{ST^n S}(s;p)=
{1\over \sqrt{2n}}e^{\pi i (3-n) \over 12}   e^{-{\pi i (p+s)^2\over 2n}}.
\fe

\bibliographystyle{JHEP}
\bibliography{Twist}

\end{document}